\documentclass[prl,twocolumn,superscriptaddress]{revtex4-1}

\usepackage{epsfig}
\usepackage{graphicx}

\usepackage{amsfonts}
\usepackage{amsmath}
\usepackage{amssymb}
\usepackage{upgreek}
\usepackage{bm}
\usepackage{dcolumn}
\usepackage{graphics}
\usepackage{latexsym}
\usepackage{rotating}
\usepackage{hyperref}
\usepackage{mathtools}
\usepackage{xspace} 
\usepackage[usenames,dvipsnames]{color}

\usepackage[caption=false]{subfig}

\newcommand\be{\begin{equation}}
\newcommand\ba{\begin{eqnarray}}
\newcommand\ee{\end{equation}}
\newcommand\ea{\end{eqnarray}}

\newcommand{\pr}{{\mbox{\tiny pr}}}
\newcommand{\rr}{{\mbox{\tiny rr}}}

\newcommand{\beq}{\begin{equation}}
\newcommand{\eeq}{\end{equation}}
\newcommand{\bes}{\begin{subequations}}
\newcommand{\ees}{\end{subequations}}
\newcommand{\beqn}{\begin{eqnarray*}}
\newcommand{\eeqn}{\end{eqnarray*}}

\newcommand{\uvec}[1]{\bm{\hat{#1}}}

\newcommand{\sub}[1]{_{\text{#1}}}

\newcommand{\ord}[1]{\mathcal{O} \left( #1 \right)}
\newcommand{\precav}[1]{\left\langle #1 \right\rangle\sub{\pr}}

\begin{document}
\renewcommand{\thefigure}{\arabic{figure}}
\setcounter{figure}{0}

\bibliographystyle{apsrev}

\title{Analytic Gravitational Waveforms for Generic Precessing Binary Inspirals}

\author{Katerina Chatziioannou}
\affiliation{eXtreme Gravity Institute, Department of Physics, Montana State University, Bozeman, Montana 59717, USA}
\author{Antoine Klein}
\affiliation{Department of Physics and Astronomy, The University of Mississippi, 
University, MS 38677, USA}
\affiliation{CENTRA, Departamento de F\'isica, Instituto Superior
T\'ecnico, Universidade de Lisboa, Avenida Rovisco Pais 1,
1049 Lisboa, Portugal}
\author{Neil Cornish}
\affiliation{eXtreme Gravity Institute, Department of Physics, Montana State University, Bozeman, Montana 59717, USA}
\author{Nicol\'as Yunes}
\affiliation{eXtreme Gravity Institute, Department of Physics, Montana State University, Bozeman, Montana 59717, USA}

\begin{abstract}
Binary systems of two compact objects circularize and spiral toward each other via the emission of gravitational waves. The coupling of the
spins of each object with the orbital angular momentum causes the orbital plane to precess, which leads to modulation of the gravitational wave
signal. Until now, generating frequency-domain waveforms for fully precessing systems for use in gravitational wave data analysis
meant numerically integrating the equations of motion, then Fourier transforming the result, which is very computationally intensive
for systems that complete hundreds or thousands of cycles in the sensitive band of a detector.  Previously, analytic solutions were
only available for certain special cases or for simplified models. 
Here we describe the construction of closed-form, frequency-domain waveforms for
fully precessing, quasicircular binary inspirals.
\end{abstract}

\maketitle

The recent detection of gravitational waves from a binary black hole merger by LIGO~\cite{Abbott:2016blz}, with a signal that is in accordance with the predictions of Einstein's
theory~\cite{TheLIGOScientific:2016src,Yunes:2016jcc, Abbott:2016apu}, is a triumph of
engineering and theoretical physics. The GW150914 signal provided our first observational encounter with strong field, dynamical gravity, as well as a chance to compare
predictions with data. Efforts to model the orbital evolution of such binary systems and their gravitational wave emission have been ongoing for a century~\cite{Blanchet:2014}. 
The nonlinearity of Einstein's field equations greatly complicates the solution to the gravitational two-body problem, with the fixed elliptical orbits of Newton's theory
replaced by orbits that tilt and precess as the bodies spiral inward and 
eventually merge. 

While complete solutions to the two-body problem in general relativity are only known numerically, accurate approximations are available for describing the early inspiral where the orbital velocity $v$ is small
compared to the speed of light. These post-Newtonian (PN) equations of motion are known completely to ${\cal{O}}(v^6)$ and partially to ${\cal{O}}(v^7)$~\cite{Blanchet:2014}. The
effective-one-body (EOB) formalism~\cite{Buonanno99}, along with calibration against solutions from numerical
relativity~\cite{Buonanno:2009qa,PhysRevLett.106.241101,Pan:2013rra,PhysRevD.89.061502,PhysRevD.82.064016, PhysRevD.93.044007}, have been
able to extend the analytic description through merger and ringdown for binary black holes. Solving the PN equations of motion analytically is challenging, especially when the orbits are eccentric or the bodies are
spinning. While solving the PN equations numerically is far less intensive than solving the full Einstein equations, it can add days or weeks to Bayesian parameter
estimation studies~\cite{Veitch:2014wba, Farr:2015lna}.  Moreover, since most analyses are performed in the frequency domain, we seek closed-form solutions that can be computed directly in frequency. For nonprecessing
systems this can be done using the stationary phase approximation, but this approximation fails for precessing systems.

Closed-form, analytic waveform models for spin-precessing systems currently exist for several special cases. The first are for systems where only one object is
spinning~\cite{Apostolatos:1994mx}. The ensuing motion is~\emph{simple precession} and the resulting waveform is ideal for
black hole - neutron star (BHNS) systems~\cite{Lundgren:2013jla}. Related to these are waveforms described 
by effective spin parameters that provide good matches to fully precessing waveforms~\cite{PhysRevD.70.104003}.
The effective spin approach has been used to produce approximate analytic waveforms describing the full inspiral, merger and ringdown of spinning black hole binaries~\cite{Hannam:2013oca, Schmidt:2014iyl}.
Analytic solutions have also been found for nearly aligned~\cite{Klein:2013qda} and slowly spinning~\cite{Chatziioannou:2013dza} systems.
The latter are accurate representations of neutron star - neutron star (NSNS) inspirals, both for detection and parameter estimation~\cite{Chatziioannou:2014bma,Chatziioannou:2014coa,Chatziioannou:2015uea}. 

Here we describe the construction of accurate, closed-form, frequency-domain waveforms for fully precessing, quasicircular PN inspirals. The solution utilizes three
main elements: the recently discovered reduction to quadratures for the conservative precessional dynamics~\cite{Kesden:2014sla}, multiple scale analysis (MSA) to exploit
the natural separation of time scales of the PN dynamics, and the shifted uniform asymptotic (SUA) method for
performing Fourier transforms of waveforms with caustics~\cite{Klein:2014bua}. For most
systems, the waveforms accurately match those found by numerically evolving the equations of motion and Fourier transforming the gravitational wave signal. The
minority that fail can be caught in advance and computed numerically. 

The PN expansion naturally introduces a separation of time scales: Newtonian 
dynamics at ${\cal{O}}(v^0)$, the first relativistic effects such as periastron 
precession at ${\cal{O}}(v^2)$, spin-orbit coupling at ${\cal{O}}(v^3)$, spin-spin coupling at ${\cal{O}}(v^4)$, orbital 
decay at order ${\cal{O}}(v^5)$, and so on.  Ignoring dissipation, the precession equations 
for circular orbits can be orbit averaged to yield a closed set of nine coupled, 
first-order ordinary differential equations for the spin angular momenta of the 
two bodies $\bm{S}_1$, $\bm{S}_2$ and the orbital
angular momentum $\bm{ L}$. These equations admit seven conserved quantities, $\{ S_1,S_2,L,\bm{J},\xi \}$, where $\bm{J} = \bm{ L} + \bm{S}_1 +\bm{S}_2$ is the total angular
momentum, $S_{1}, S_{2}$ and $L$ are the magnitudes of the angular momenta three-vectors, and $\xi$ is the mass-weighted effective spin
\be
\xi \equiv (1+q)\bm{S}_1 \cdot \bm{\hat{L}}  +(1+q^{-1})\bm{S}_2 \cdot \bm{\hat{L}},
\ee
where $q=m_2/m_1$ is the mass ratio. Kesden {\it et al.}~\cite{Kesden:2014sla} 
showed that by working in a noninertial, co-precessing frame of reference, the 
motion could be reduced to quadratures
in terms of the squared spin magnitude $S^2 = (\bm{S}_1 + \bm{S}_2)^2$:
\be
\left( \frac{d S^2}{dt} \right)^2 = - A^2 \left(S^6 + B S^4 + C S^2 
+D \right).\label{dSdt-prel}
\ee
where the constants $\{A,B,C,D\}$ are given in terms of the seven conserved quantities. Rather than integrate this equation numerically~\cite{Kesden:2014sla}, we were able to find
a closed-form solution in terms of Jacobi elliptic functions:
\be
S^2 = S_+^2 + (S_-^2 - S_+^2) \, \rm{sn}^2({\psi,m}) \label{Ssq-sol}
\ee
where $\rm{sn}$ is the sine-like Jacobi elliptic function with modulus $m =  (S_+^2-S_-^2)/(S_+^2-S_3^2)$ and phase $\psi = (A/2) \sqrt{S_+^2 - S_3^2}\, t$, where
$\{S_+^2, S_-^2, S_3^2\}$ are the roots of the cubic that appears on the right-hand side of Eq. (\ref{dSdt-prel}).  The solution is completed by
solving for the precession angle $\phi_z$ between $\bm{L}_\perp = \bm{ L} - (\uvec{J}\cdot \bm{ L})\uvec{J}$ and the $\uvec{x}$ direction in a coordinate system
where $\uvec{J}$ defines the $\uvec{z}$ direction and $\bm{ L}_\perp$ at some reference frequency defines the $\uvec{x}$ direction. The rate of precession $\Omega_z = \dot\phi_z$ is given by
\be
\frac{\Omega_z}{J} = a + \frac{c_0 + c_2 \,\text{sn}^2(\psi, m) + c_4 \,\text{sn}^4(\psi, m)}{d_0 + d_2\, \text{sn}^2( \psi, m ) + d_4 \,\text{sn}^4( \psi, m)},
\ee
where the constants $\{a, c_0, c_2, c_4, d_0, d_2, d_4\}$ are given in terms of the seven constants of the motion and the orbital velocity. This equation
can be integrated to give $\phi_z$ in terms of elliptic integrals. The remaining 
angles needed to specify $\bm{S}_1$, $\bm{S}_2$
and $\bm{L}$ are given in terms of $S(t)$ and the constants of the motion. This completes the construction of a closed-form, analytic solution to the conservative dynamics.

The emission of gravitational radiation causes the system to lose energy and angular momentum. Here we can use the separation between the precession time scale
$T_{\pr} \equiv |\bm{S}_1|/|\dot{\bm{S}}_1| \sim  v^{-5}$ and the radiation-reaction time scale $T_{\rr}  \equiv v/ \dot{ v} \sim  v^{-8}$ to develop a MSA
solution that incorporates dissipation. For most variables we find that the leading-order term in the MSA is sufficient.
Additional accuracy could be achieved by continuing to
higher order in the expansion. Of the original seven constants of motion, the 
spin magnitudes $\{ S_1,S_2, \xi \}$ remain constant under radiation reaction. 
While the magnitude of the total angular momentum $J$ changes as $L$ decays, the direction $\uvec{J}$ remains almost constant. This can be
established by precession averaging and PN expanding  the evolution equation $\hspace{0.15cm} \dot{\hspace{-0.15cm}{\uvec{J}}} = \dot{L}  \uvec{L}/J  - \dot{J} \uvec{J}/L$ to show that $\uvec{J}_z$ is
constant to ${\cal{O}}(v^2)$, while $\uvec{J}_{x,y}$ oscillate but exhibit no secular growth at ${\cal{O}}(v^2)$. Since the wobble in $\uvec{J}$ is very small, we are able to
neglect this variation and continue to use $\uvec{J}$ to define our coordinate system.  This preserves the geometrical framework used to solve the spin-precession equations.

The orbital angular momentum depends on the orbital velocity as $L = (m_1+m_2) \eta /v$, where $\eta = m_1 m_2/(m_1+m_2)^2$ is the symmetric mass ratio. Precession averaging and
PN expanding the evolution equation for $J$ yields $J^2 = L^2  + 2 c_1/v + c_2  + {\cal O}(v)$, where $c_1$ and $c_2$ are constants that are set by the initial 
conditions at some reference frequency. The evolution of $L$ and $J$ causes the roots $\{S_+^2, S_{-}^2, S_{3}^2\}$ to evolve on the radiation-reaction time scale: $S_{\pm}^2 = S_{\pm,0}+ \ord{v}$
and $S_{3}^2 = \ord{v^{-2}}$. The MSA solution for $S^2$ follows from  adiabatically promoting the constants in Eq.~(\ref{Ssq-sol}) to functions of time. 
To leading order, the amplitude of the oscillations in $S^2$ are constant, while the modulus grows as $m \sim v^2$; thus, the oscillations become increasingly anharmonic as the masses spiral towards
each other. The phase $\psi$ can be PN expanded and integrated: $\psi = \psi_0 - 3  g_0 (m_1-m_2)(1+ \psi_1 v +  \psi_2  v^2 +\dots)/(4 \, v^3 )$ where $\{\psi_0,g_0, \psi_1,\psi_2\}$ are constants
that depend on the masses, spins and initial conditions. Comparison with the numerical solution shows that the leading-order MSA solution for $S^2$ is very accurate, so there
is no need to continue to higher orders in the expansion. Finding a solution for $\phi_z$ to complete the derivation for precessional motion with dissipation turns out to be the most challenging step.
To compute the MSA, we introduce two time variables, the precession time $t_{\pr}$ and the radiation-reaction time $t_{\rr}$. The leading-order term in the MSA is found by precession averaging $\Omega_z$ and PN integrating:
\be
\phi_{z,-1} = \int \precav{\Omega_z}\!(t_{\rr}) \; dt_{\rr} = \int  \langle \Omega_z \rangle_{\pr} \, \frac{dv}{\dot v} \, .
\ee
While this leading-order term captures the overall secular evolution of $\phi_z$, we found that the agreement between the numerical and analytic solutions could be improved by
including the second-order term in the MSA expansion,
\be
\phi_{z,0} =  \int  \Omega_z(t_{\pr}, t_{\rr})\; dt_{\pr} - \int \precav{\Omega_z}\!(t_{\rr}) \; dt_{\pr} \, ,
\ee
which describes small oscillations in $\phi_z$ on the precession time scale.

\begin{figure}[htbp]
\includegraphics[width=\columnwidth,clip=true]{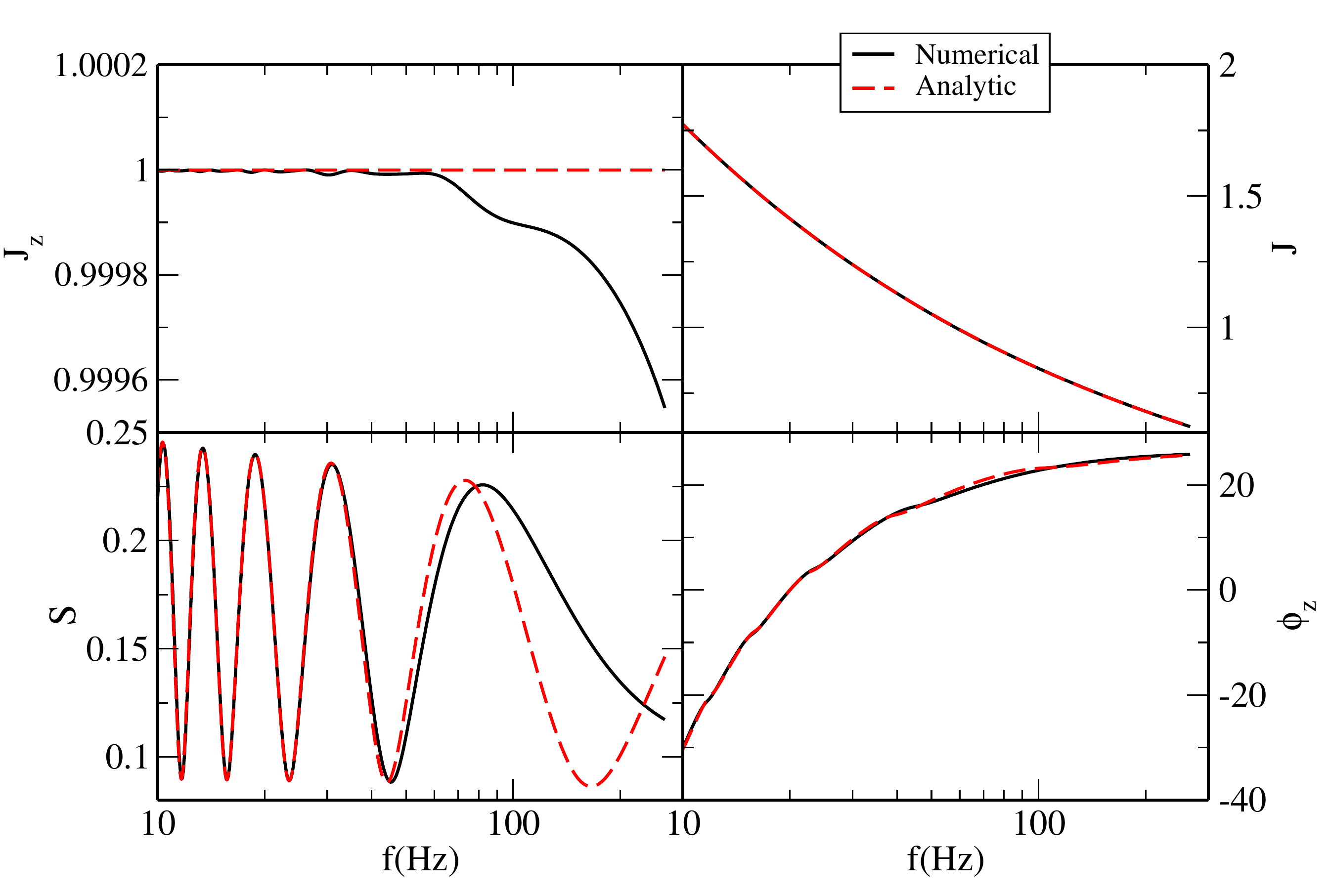}
\caption{\label{fig:comparisons} A comparison of the analytic (red, dashed lines) and numerical (black, solid lines) solutions to the PN equations of motion for a strongly precessing black hole binary. Clockwise from the
upper left we have the frame orientation $\uvec{J}_z$, the magnitude of the total angular momentum $J$, the precession angle $\phi_z$, and the magnitude of the total spin $S$.
The largest discrepancy between the solutions occurs for $S$ and can be traced to the approximate phasing $\psi$. However, as shown in Fig.~\ref{fig:hcomparisons}, the impact on the waveform is minor.}
\end{figure}

Figure~\ref{fig:comparisons} compares the analytic and numerical solutions for $\uvec{J}_z$, $J$, $S$, and $\phi_z$ for a strongly precessing black hole -  black hole (BHBH) binary with masses $(m_1, m_{2})=(10,7) M_\odot$
and spin magnitudes $(S_1,S_{2}) = (0.6 m_1^2, 0.7 m_2^2)$. The angles between the spins and angular momentum at a reference frequency of 20 Hz are $\bm{L},\bm{S} = 78^\circ$, $\bm{L},\bm{S}_1 = 120^\circ$,
and $\bm{L},\bm{S}_2 = 36^\circ$. Overall, the agreement is excellent. The largest discrepancy occurs in the spin magnitude, and it is due to the difference between the numerical and
analytic solutions for the phase $\psi$. It is possible to improve the agreement by continuing the PN expansion of $\psi$ to higher order, but this is unnecessary since the dephasing in $S$ has little
impact on the waveforms. Note that the waveform was truncated at $r = 6(m_1+m_2)$, and does not extend to cover the merger and ringdown portions of the full signal.

With the analytic solution for the orbital motion in hand, the next step is to produce the gravitational waveforms in the frequency domain using the SUA transform. The gravitational wave signal emitted by 
a binary system in general relativity as observed in an interferometric detector is $h(t) = F_+h_+ +  F_\times h_\times$,
 where $(F_+,F_\times)$ are the antenna pattern functions and $(h_+,h_\times)$ are the two polarization states of the gravitational wave signal.
 The polarization states for a source located in the $\uvec{N}$ direction can be decomposed into a spin-weighted spherical 
harmonic basis~\cite{Arun:2008kb}
\begin{align}
 h_+ - i h_\times &= \sum_{l \geq 2} \sum_{m = -l}^l H_{lm}(
\theta_s, \phi_s) e^{- i m \Phi},
\end{align}
where $\Phi  =  \phi\sub{orb} - 3  v^3 (2 - \eta  v^2 ) \ln v\,,$ $\phi\sub{orb}$ is the orbital phase, $(\theta_s,\phi_s)$ are the spherical angles of $\uvec{N}$ in a frame where $\uvec{J}$ is along 
the $z$ axis and 
\begin{align}
 H_{lm} &= h^{lm} \sum_{m' = -l}^l D^l_{m',m} (\phi_z, \theta_L, \zeta)
{}_{-2}Y_{lm'} (\theta_s, \phi_s),
\end{align}
where the amplitudes $h^{lm}$ can be found in Ref.~\cite{Blanchet:2014}, 
$D^l_{m,m'}$ are the Wigner 
D matrices, ${}_s Y_{lm}$ are the spin-weighted spherical harmonics, the angles 
$\theta_L$ and $\phi_z$ are the spherical angles of $\uvec{L}$ in 
the same frame as $\theta_s$ and $\phi_s$ are defined, and $\zeta$ satisfies 
$\dot{\zeta} =  \dot{\phi_z} \cos{\theta_L}$. In order to solve for $\zeta$, we employ the same MSA techniques as for $\phi_z$.

To compute the Fourier transform of $h$, we use the SUA method devised in Ref.~\cite{Klein:2014bua} and write
\begin{align}
 \tilde{h}(f) &= \sqrt{2\pi} \sum_{m \geq 1} 
 T_m e^{i(2 \pi f t_m - m \Phi - \pi/4)} \nonumber\\
&\times \sum_{l 
\geq 2} \sum_{k=-k\sub{max}}^{k\sub{max}} \frac{a_{k,k\sub{max}}}{2 - 
\delta_{k,0}} \mathcal{H}_{lm}( t_m + k T_m )\label{SUAGW},
\end{align}
where $t_m$ and $T_m$ are defined implicitly by $2 \pi f = m \dot{\Phi}(t_m)$, $T_m = (m \ddot{\Phi}(t_m))^{-1/2}$, and
\be
 \mathcal{H}_{lm} = \frac{1}{2} (F_+ + i F_\times)H_{lm} + \frac{1}{2} (F_+ - i F_\times) H_{l, -m} 
 \ee
with the constants $a_{k, k\sub{max}}$ satisfying the linear system
\begin{align}
 \frac{(-i)^p}{2^p p!} &= \sum_{k=0}^{k\sub{max}} a_{k, k\sub{max}} 
\frac{k^{2p}}{(2p)!},
\end{align}
for $p \in \{ 0, \ldots, k\sub{max} \}$.
For a static  detector, $\mathcal{H}_{lm}$ depends on frequency
only through $\phi_z$, $\theta_L$, and $\zeta$. As shown in 
Ref.~\cite{Klein:2014bua}, setting $k\sub{max}=3$ is sufficient
to accurately match the numerical Fourier transform.

\begin{figure}[htbp]
\includegraphics[width=\columnwidth,clip=true]{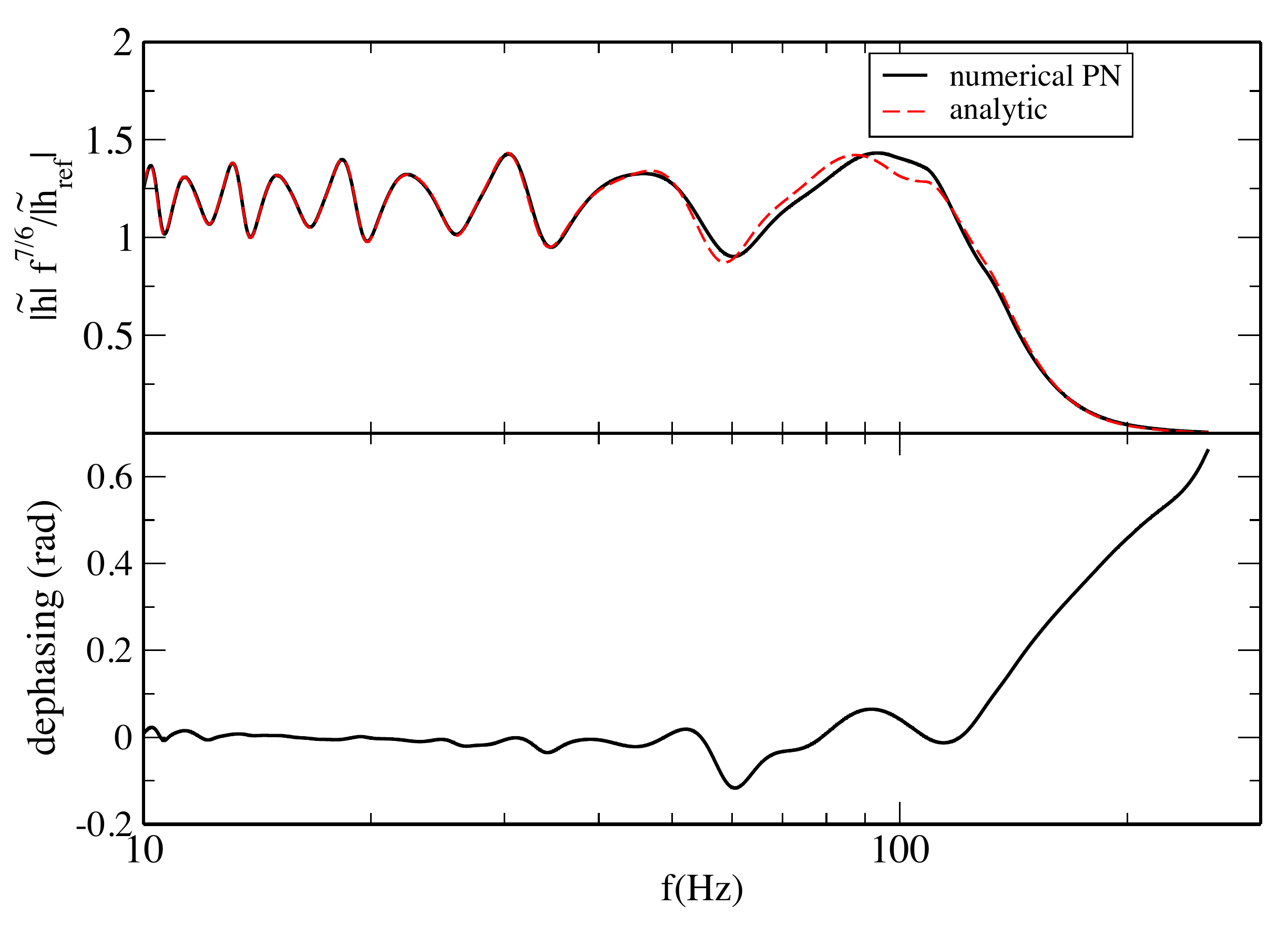}
\caption{\label{fig:hcomparisons} A comparison of the analytic and numerical solutions for the Fourier amplitude (upper panel) and Fourier phase (lower panel) of the observed gravitational wave signal
for a strongly precessing black hole binary. The amplitudes are scaled by a reference amplitude at 20 Hz, $|\tilde{h}_{\rm{ref}}| \equiv (20\,  {\rm Hz})^{7/6} |\tilde{h}(f=20)|$, and multiplied by $f^{7/6}$ to account for the dominant secular evolution.}
\end{figure}

Figure~\ref{fig:hcomparisons} compares the numerical and analytic solutions for the amplitude and phase of the gravitational waveform $h$ produced by the system shown in Fig.~\ref{fig:comparisons}, observed edge-on and located on the detector plane at the initial reference time. 
There is good agreement across the band, with the dephasing at high frequencies coming from errors in the PN integrated phase $\psi$. The discrepancy has little effect on the overlap between the waveforms, which
we measure in terms of the~\emph{faithfullness}, defined as
\begin{align}
F &\equiv\underset{ t_c, \phi_c}{\text{max }} 
\frac{\left(h_{1}\left|\right.h_{2}\right)}{\sqrt{\left(h_{1}\left|\right.h_{1}\right)
\left(h_{2}\left|\right.h_{2}\right)}}.
\label{match}
\end{align}
where $t_c$ and $\phi_c$ are the merger time and phase, and $(a\vert b)$ denotes the usual noise-weighted inner product. Using the aLIGO design zero-detuning, high-power noise spectral density~\cite{AdvLIGO-noise},
the faithfulness of the analytic waveform shown in Fig.~\ref{fig:hcomparisons}  is $F=0.9997$. 

\begin{figure}[htbp]
\includegraphics[width=\columnwidth,clip=true]{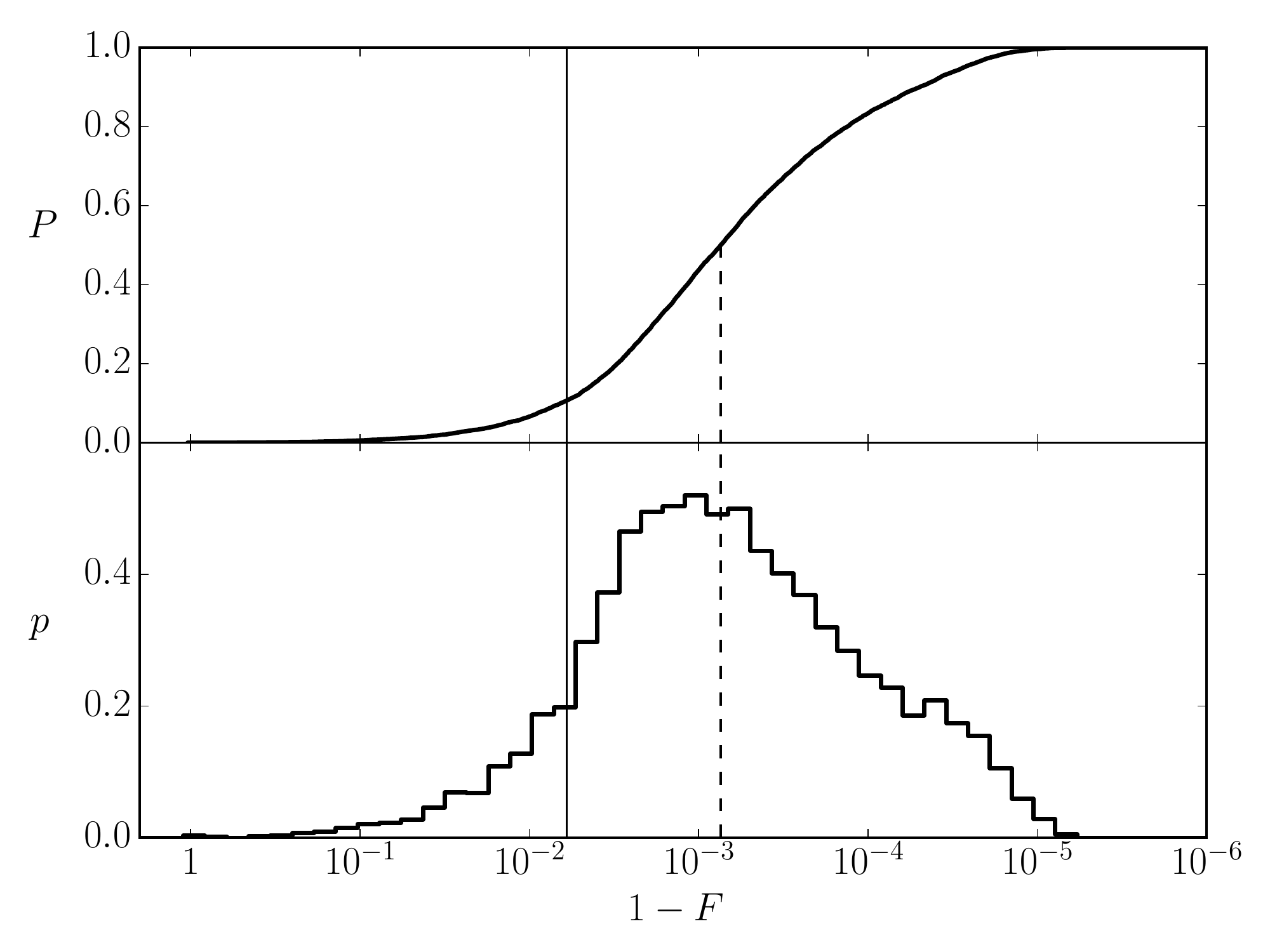}
\caption{\label{fig:faith} Cumulative (upper panel) and fractional (lower panel) distribution of the unfaithfulness, $1-F$, of the analytic waveforms for a sample of 10,000 precessing
binary black hole systems. While the agreement between the
numerical and analytic PN waveforms is excellent for the majority of systems, 
there exists a small subset that produce unacceptably high mismatches  above a nominal threshold of $F=0.994$ indicated by the solid vertical line.
The dashed vertical line marks the median unfaithfulness, which corresponds to $1-F=7.4 \times 10^{-4}$.}
\end{figure}

We tested the analytic waveforms for a wider range of signals through a 
Montwatch moonlight onlinee Carlo study that covered NSNS, NSBH and BHBH 
binaries. Here we focus on the BHBH systems since
they have the most complicated precessional dynamics, and their inspiral is not described by existing analytic methods.
The Monte Carlo study drew 10,000 systems with masses drawn uniformly in logarithm between $[2.5,20]M_{\odot}$, and dimensionless spin magnitudes $S_i/m_i^2$ drawn uniformly in
$[0,1]$. The initial directions of the unit vectors $\{\uvec{L}, \uvec{S}_1, \uvec{S}_2\}$ and the sky location $\uvec{N}$ were drawn randomly on the sphere. Figure~\ref{fig:faith} shows the
distribution of the unfaithfulness $1-F$ for this sample. We demand that the systematic errors introduced by waveform modeling errors are smaller than the statistical errors. This requires choosing
a reference SNR for the systems of interest, as the statistical errors scale with the SNR while the systematic errors are SNR-independent~\cite{Lindblom:2008cm}. It can be shown that the
expected value for the faithfulness due to statistical errors in the intrinsic parameters is given by $F = 1 - D_{\rm in}/(2\, {\rm SNR}^2)$, where $D_{\rm in}$ are the number of intrinsic parameters.
Choosing a reference ${\rm SNR}=25$ and using that $D_{\rm in} = 8$, we obtain a 
nominal accuracy threshold of $F=0.994$. We found that 10.7\% of BHBH systems fell outside of this accuracy requirement (for NSNS binaries, the fraction was 0.3\%, and for BHNS binaries 1.6\%).
We found that systems with very low overlaps $F < 0.97$ fell into three categories. The first category includes systems with total angular moment $\bm{L}$ and orbital angular momenta $\bm{S}$ that pass through
near-anti-alignment during the evolution of the orbit, which leads to a problem with our coordinate system which is defined by $\bm{J} = \bm{L} + \bm{S}$ and ${\bm L} \times \bm{J}$. In particular,
the $\phi_z$ coordinate becomes ill defined when ${\bf S}$ and ${\bf L}$ are parallel. So long as the alignment is not perfect, the numerical solution proceeds smoothly, while the analytic MSA PN expansion of $\phi_z$ has coefficients in the velocity expansion that diverge. The second category of troublesome cases are nearly edge-on systems, ($\bm{L}\cdot\bm{N}\sim0$), a configuration which maximizes the effects of precession on the waveforms, and amplifies any small inaccuracies in the analytic solution.  The third category of bad systems were found to undergo {\em transitional precession}~\cite{Apostolatos:1994mx}. We found
that the overlaps could be improved in all cases by going to higher order in the MSA and the PN integration, but to fully solve the problem, one will likely need a change of coordinates.
In data analysis applications, the troublesome systems can be caught in advance and other, slower methods, such as the
numerical SUA~\cite{Klein:2014bua} can be used to generate the waveforms. By a change of coordinates, or by some other means, it should be possible to modify the
$\phi_z$ solution so that it can handle spin-orbit anti-alignment and transitional precession. Extending the analytic solution to higher order will improve the fitting factor for edge-on systems.
We leave these extensions to future work.

In summary, we have described the derivation of the first closed-form, frequency-domain waveforms for fully precessing compact binary inspiral. Complete details will be provided in a longer follow-up paper.
The method described here can be extended to cover the full inspiral, merger and ringdown stages of a black hole merger using EOB or phenomenological waveforms. In particular, our work allows for the
development of fully precessing variants of the effective-spin ``PhenomP'' waveforms~\cite{Hannam:2013oca,Schmidt:2014iyl}. 
The new waveforms are typically much faster to compute than traditional numerical, time-domain implementations - up to 3 orders of magnitude faster for NS-NS binaries starting from 10 Hz.
The analytic solution also provides additional physical insight into the dynamics.

\section*{Acknowledgments}
We would like to thank Emanuele Berti, Mike Kesden, Sylvain Marsat, and Frank Ohme for helpful discussions and suggestions.
K.~C. acknowledges support from the Onassis Foundation. N.~Y. acknowledges 
support from NSF CAREER Grant No. PHY-1250636. N.~C. 
acknowledges support from the NSF Award PHY-1306702.
N.~C. and N.~Y. acknowledge support from NASA Grant No. 
NNX16AB98G. A.~K. is supported by NSF CAREER Grant No. PHY-1055103, and by 
Funda{\c{c}}{\~{a}}o para a Ci{\^{e}}ncia e a Tecnologia (FCT) 
contract IF/00797/2014/CP1214/CT0012 under the IF2014 
Programme.

\bibliography{letter}

\end{document}